\documentclass[9pt,twocolumn,twoside]{osajnl}

\journal{josaa} 

\setboolean{shortarticle}{false}

\usepackage{multirow}
\definecolor{red}{rgb}{0.9, 0.1, 0.1}

\usepackage{lineno}
\usepackage{float}
\usepackage{hyperref}

\title{Modal decomposition of complex optical fields using convolutional neural networks}

\author[1,2,*]{Mitchell G. Schiworski}
\author[1,2]{Daniel D. Brown}
\author[1,2]{David J. Ottaway}

\affil[1]{Department of Physics and The Institute of Photonics and Advanced Sensing (IPAS), University of Adelaide, SA, 5005, Australia}
\affil[2]{OzGrav, Australian Research Council Centre of Excellence for Gravitational Wave Discovery}

\affil[*]{Corresponding author: mitchell.schiworski@adelaide.edu.au}

\begin{abstract}
Recent studies have shown convolutional neural networks (CNNs) can be trained to perform modal decomposition using intensity images of optical fields. A fundamental limitation of these techniques is that the modal phases can not be uniquely calculated using a single intensity image. The knowledge of modal phases is crucial for wavefront sensing, alignment and mode matching applications. Heterodyne imaging techniques can provide images of the transverse complex amplitude \& phase profile of laser beams at high resolutions and frame rates.
In this work we train a CNN to perform modal decomposition using simulated heterodyne images, allowing the complete modal phases to be predicted. This is to our knowledge the first machine learning decomposition scheme to utilize complex phase information to perform modal decomposition. We compare our network with a traditional overlap integral \& center-of-mass centering algorithm and show that it is both less sensitive to beam centering and on average more accurate.
\end{abstract}

\setboolean{displaycopyright}{true}

\begin{document}

\maketitle

\section{Introduction}
Heterodyne imaging techniques are used in a variety of high precision optical experiments. An example of this is in gravitational wave interferometers, where heterodyne techniques are used throughout to sense various length \& misalignment degrees of freedom within the interferometer \cite{barsotti_alignment_2010,aso_length_2012,allocca_interferometer_2020}. This is achieved by demodulating the beat of the various radio-frequency sideband fields with the main carrier field measured on single \& quadrant element photo-diodes. This technique provides a high bandwidth sensing solution, but is limited to low-resolution segmented photodiodes.
\paragraph{}
Future upgrades to gravitational wave interferometers will require using higher power lasers \cite{and_advanced_2010,acernese_advanced_2014} and increasing the amount of squeezed light injected into the interferometer \cite{barsotti_squeezed_2018} to reduce quantum noise effects and improve sensitivity. This will place a greater stress and importance on current alignment \cite{barsotti_alignment_2010} and mode mismatch control systems \cite{perreca_analysis_2020}. Maximising the sensitivity benefits of these upgrades will require higher spatial resolution heterodyne techniques to allow the diagnosis of the control systems by investigating higher order mode content in the sideband fields. This has previously been achieved with the use of phase cameras \cite{goda_frequency-resolving_2004,gretarsson_effects_2007,betzwieser_analysis_2008,agatsuma_high-performance_2019}. Higher spatial resolution measurements represent a greater challenge in the fast processing and analysis of the data produced. Decomposing beams into orthogonal basis modes is an effective way of compressing high spatial resolution images and is useful for the analysis of the interferometer state. Traditional decomposition methods that utilize overlap integrals can be cumbersome to apply and difficult for real-time analysis at higher resolutions. Machine learning techniques may be able to overcome these difficulties.
\paragraph{}
Very recently, work has been done using CNNs as tools to analyse the mode content of both free space and fibre lasers using intensity images. In \cite{doster_machine_2017} a CNN was trained to successfully demultiplex orbital angular momentum modes of free space lasers and in \cite{an_learning_2019} a CNN was trained to perform mode decomposition of few mode fibres. In terms of free space Hermite-Gaussian (HG) mode analysis, \cite{hofer_hermitegaussian_2019} trained a CNN to classify images of single HG modes to a high accuracy and \cite{an_fast_2020} trained a CNN to perform fast HG modal decomposition. A fundamental limitation of intensity image based modal decomposition is the sign ambiguity in calculating the modal phases due to the lack of phase information in intensity images. As such the techniques described in \cite{an_learning_2019} and \cite{an_fast_2020} allow the prediction of the modal phases only up to a sign ambiguity. This limitation can be overcome by using two intensity images - one each in the near and far field \cite{cutolo_transverse_1995}, but still involves a complex optimization process to calculate the mode coefficients.
\paragraph{}
Knowledge of the sign of the modal phase is important in the analysis of gravitational wave interferometer sidebands. Particularly because it distinguishes first order modes generated from transverse \& angular beam misalignments into an optical cavity as well as second order modes generated from an incorrect beam waist size \& axial position mismatches with respect to the cavity \cite{anderson_alignment_1984}.
\paragraph{}
Heterodyne imaging techniques like the phase camera measure both the transverse amplitude and phase of a beam which allows the modal decomposition to be performed without any ambiguity in the measured mode amplitudes or phases. The process is not straightforward however as one needs to account for the effects of the reference field used in the imaging process which spatially envelopes the measured field. Under experimental conditions the beam is subject to wandering which results in images where the actual beam center can move. This is problematic for mode decomposition because if not accounted for will result in the measured mode coefficients changing as the beam wanders. The traditional computational approach to mode decomposition of phase camera images involves multiple computational steps which is problematic when a real time analysis of streaming phase camera images is desired. Training a CNN to receive phase camera images and perform the decomposition process allows the reference field unwrapping, beam centering and decomposition steps to be combined into a single CNN. Deploying the CNN on GPUs or field-programmable gate arrays (FPGAs) allows mode decomposition at speeds suitable for real-time analysis and implementation into adaptive wavefront control loops.
\paragraph{}
The process of training a CNN to perform mode decomposition on phase camera images is a different problem to doing so with intensity images alone. The phase information provided by the phase camera requires the input to the network to be complex, resulting in a depth-wise increase in dimensionality. The ability to predict the modal phases without a sign ambiguity also changes the activation functions needed for the final output layers of the network. The non-centered beam images and unwrapping of the reference field increase the complexity of the network's task and parameter space that needs to be sampled in the training process. In this paper we explore the use of a convolutional neural network (CNN) to perform HG mode decomposition on simulated phase camera images and importantly compare the results with a traditional overlap integral \& center of mass approach.
\paragraph{}
This paper is structured as follows: Section 2 gives a brief introduction to HG mode decomposition, and the working principle and applications of phase cameras are described in Section 3. Section 4 outlines the overlap integral decomposition algorithm which is used to compare the effectiveness of the decomposition network. Section 5 describes the structure and training process of the mode decomposition network. Finally, in Section 6 the decomposition network and integral algorithm are compared across a testing dataset of simulated phase camera images. This test is performed at increasing decomposition orders and with images at 64x64 \& 128x128 resolution.
\section{Hermite-Gaussian mode decomposition}
\subsection{Hermite-Gaussian modes}
The HG modes are a complete and orthonormal set of solutions to the free-space paraxial wave equation. They are of mathematical and physical convenience in the description of laser fields inside resonators which exhibit rectangular symmetry. Assuming no astigmatism (i.e $\Tilde{q}_x=\Tilde{q}_y$), the electric field of the n$^{th}$ and m$^{th}$ order HG mode propagating in the $\hat{z}$ direction is \cite{kogelnik_laser_1966}
\begin{equation}
\label{HG_mode_eqn}
    \begin{aligned}
        & \text{HG}_{nm}(x,y,z,\Tilde{q}) =  \sqrt{ \frac{2}{\pi} }  \left( \frac{1}{2^{m+n}m!n!} \right)^{1/2} \frac{1}{\omega(z)}\\
        & \times H_n\left( \frac{\sqrt{2}}{\omega(z)}x \right) H_m\left( \frac{\sqrt{2}}{\omega(z)}y \right)\\
        & \times \exp \left[ -i\frac{k(x^2 + y^2)}{2\Tilde{q}(z)} + i(1+n+m)\psi(\Tilde{q}) \right], \\
        & \Tilde{q}(z) = z + i\frac{\pi\omega^2_0}{\lambda} \text{, or equally: } \frac{1}{\Tilde{q}(z)} = \frac{1}{R(z)}-i\frac{\lambda}{\pi\omega(z)}\\
        & \psi(\Tilde{q}) = \tan^{-1}\left( \frac{\operatorname{Re}(\Tilde{q})}{\operatorname{Im}(\Tilde{q})} \right)
    \end{aligned}
\end{equation}
where $H_i$ is the $i^{th}$ order Hermite polynomial, $\Tilde{q}(z)$ is the complex beam parameter, $\omega_0$ is the beam waist, $\lambda$ is the beam wavelength, $R(z)$ is the radius of curvature of the beam, $\omega(z)$ is the spot size of the beam and $\psi(\Tilde{q})$ is the Gouy phase.
\paragraph{}
A set of HG modes is described by their complex beam parameter, $\Tilde{q}(z)$, which forms the basis of the set. The imaginary component of $\Tilde{q}(z)$ is related to the ratio of the transverse extent of the beam at its focus, $\omega_0$, and it's wavelength, $\lambda$. The real part of $\Tilde{q}(z)$ is the axial distance from the beam focus, with positive/negative values of $z$ indicating a diverging/converging beam. \cite{siegman_lasers_1986}
\paragraph{}
The lowest order HG$_{00}$ mode, also known as the Gaussian mode due to its Gaussian transverse intensity profile, is the fundamental mode of most free space lasers and often the desired mode of propagation inside optical systems. The modes of propagation of optical systems can be very well approximated by the HG modes and hence it is useful to describe the optical field distribution in these systems in terms of HG modes.
\begin{figure}[h]
    \centering
    \includegraphics[width=0.7\linewidth]{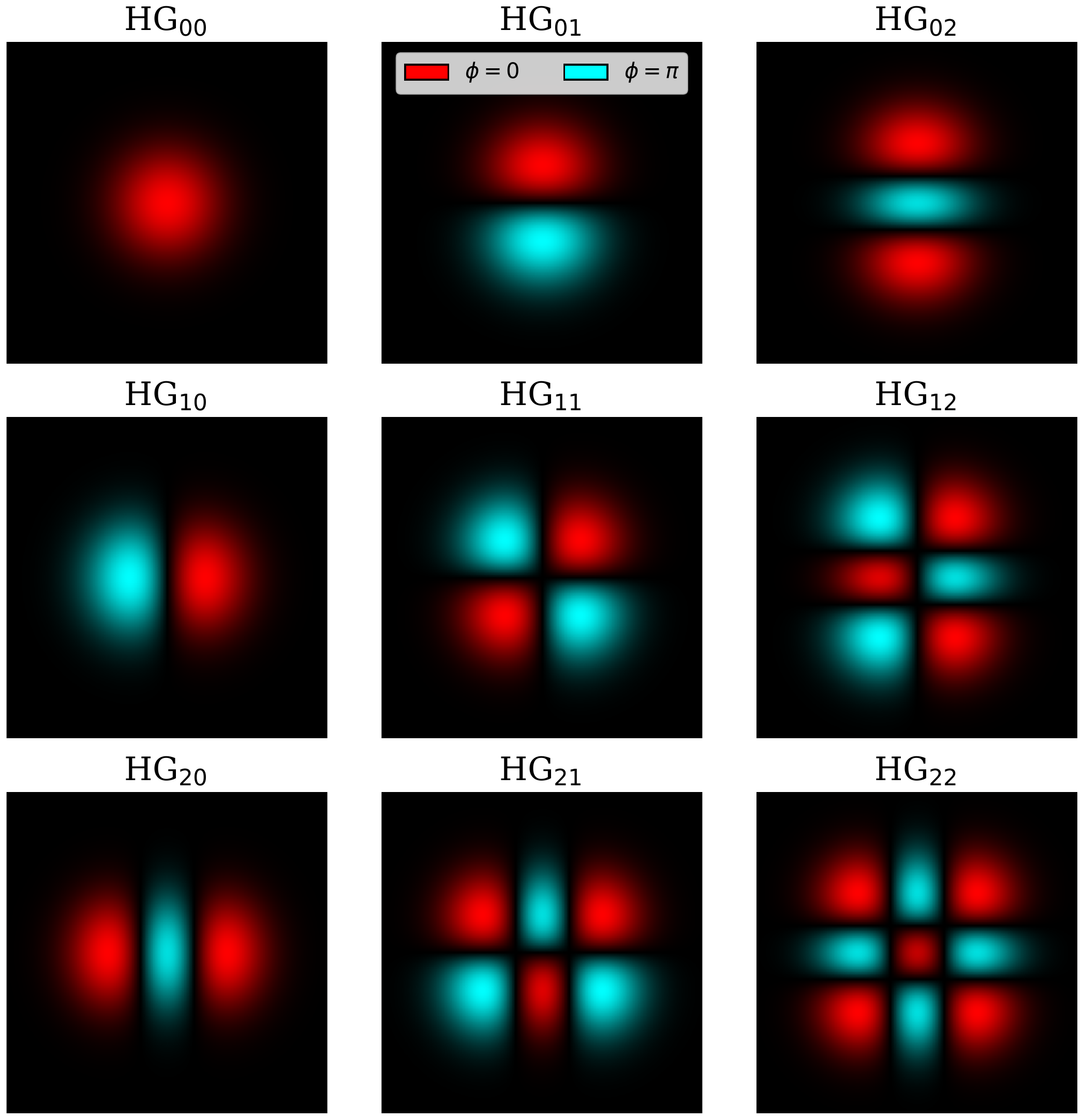}
    \caption{The transverse amplitude and phase profiles of the first 9 Hermite-Gaussian modes evaluated at the beam waist plane ($z=0$). Here the phase information is represented with colour to highlight the $\pi$ phase shift between different lobes. The beam wavefront is exactly planar only at the waist, at other planes the wavefront curvature and Gouy phase will effect the transverse phase profile.}
    \label{fig:hg_modes}
\end{figure}

\subsection{Mode decomposition}
Since the HG modes form a complete orthonormal basis, any arbitrary electric field $U(x,y,z)$ can be represented as a superposition of these modes with a particular $\Tilde{q}$ parameter \cite{siegman_lasers_1986}
\begin{equation}
    \label{mode_decomp_eqn}
    \begin{split}
        U(x,y,z) &= \sum_{n=0}^{\infty}\sum_{m=0}^{\infty} c_{nm}\text{HG}_{nm}(x,y,\Tilde{q})\exp(-ikz)\\
        & c_{nm} = r_{nm}\exp[i\theta_{nm}]\\
    \end{split}
\end{equation}
\begin{equation*}
    \text{for a normalised field:}\sum_{n=0}^{\infty}\sum_{m=0}^{\infty} |c_{nm}|^2 = 1
\end{equation*}
where $c_{nm}$ are the complex mode coefficients and $r_{nm}$ \& $\theta_{nm}$ are the mode amplitudes \& phases.

\paragraph{}
HG mode decomposition is the process of calculating these mode coefficients, which can be done by exploiting the orthogonality of the HG modes and integrating across the transverse plane at some axial location, $z_0$ \cite{siegman_lasers_1986}
\begin{equation}
    \label{mode_coeff_integral_eqn}
    c_{nm} = \int_{-\infty}^{\infty}\int_{-\infty}^{\infty} U(x,y,z_0)HG^*_{nm}(x,y,\Tilde{q})dx\, dy.
\end{equation}
The measured $c_{nm}$ are dependant on the choice of basis (or $\Tilde{q}$ parameter) for decomposition. There is no \emph{unique} $\Tilde{q}$ that describes a particular electric field because each set of HG modes is complete and hence \emph{any} $\Tilde{q}$ can be used to exactly describe \emph{any} field. As such comparisons between sets of mode coefficients must be done in the same basis so that observed changes in mode content are not due to different basis projections.
\paragraph{}
The mode coefficients measured from an image are also dependant on other parameters unrelated to the beam shape. The main being the choice of origin in the transverse plane ($x_0, y_0$) which can change in an experimental setting as the beam position wanders. The integration limits and window shape can also have an effect and the amount of modes required to adequately describe a beam varies. A general rule of thumb is that systems containing a predominantly Gaussian beam with misalignment/mismatch less than 10\% require the integration limits extend at least 3 times the spot size in each direction and a maximum order of $n+m \leq 6$ to sufficiently describe the beam \cite{siegman_lasers_1986}.
\section{Phase Cameras}
\begin{figure}[h]
    \centering
    \includegraphics[width=\linewidth]{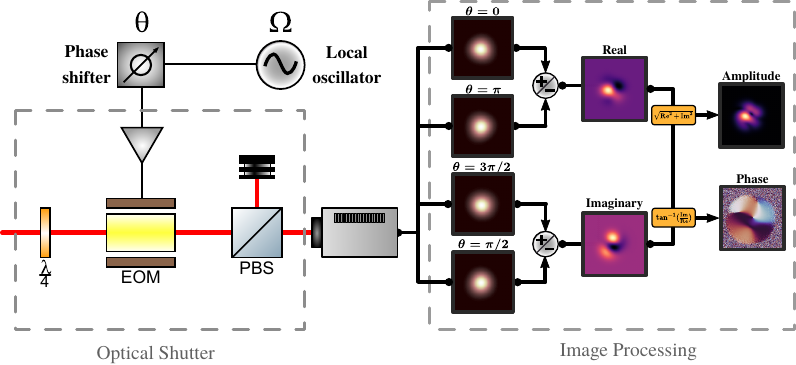}
    \caption{A simplified diagram of the optical lock-in phase camera \cite{cao_optical_2020}. The quarter-wave plate, electro-optic modulator (EOM) and polarising beam splitter (PBS) combine to act as an amplitude modulator with a 50\% transmission when no voltage is applied to the EOM. The process is analogous to electrical demodulation of a single photo-diode where instead the mixing with the local oscillator is performed optically by this amplitude modulator. The camera acts as a low-pass filter and each CCD acts as a single demodulated photo-diode. The complex amplitude of the field at the local oscillator frequency is reconstructed via taking images at four different demodulation phases $\theta = [0,\pi,3\pi/2,\pi/2]$ and combining them as shown above.}
    \label{fig:phase_cam_simplified}
\end{figure}

Phase cameras work via demodulating the heterodyne beat of a signal and reference field at multiple transverse positions to recreate the amplitude and phase profile of the beat field. This has been achieved by mechanically scanning the beam over a single photo-diode \cite{goda_frequency-resolving_2004,agatsuma_high-performance_2019} and more recently using optical demodulation schemes \cite{cao_optical_2020,panigrahi_all-optical_2020}. Figure \ref{fig:phase_cam_simplified} describes the operation of the optical lock-in phase camera design. Other techniques utilize devices which can perform on-board demodulation where each pixel is paired with an integrated demodulation circuit \cite{patel_widefield_2011}, or devices with synchronized camera shutters \cite{cervantes_real-time_2007}.
\paragraph{}
The self referencing or external referencing configurations of a phase camera refer to whether the beat field being imaged is that of a carrier and sideband (self referencing) or of two separate frequency shifted fields (external referencing). In the self referencing case the wavefront curvature of the carrier and sideband field cancel each other out, meaning that measurements of the exact wavefront must be done with an external reference field with known phase. Since the curvature of the wavefront is related to the distance from the beam waist and not the modal content this is not an issue for modal decomposition and has the advantage of removing the irrelevant portion of the phase information.
\paragraph{}
This work focuses on images from the optical lock-in type camera described in Figure \ref{fig:phase_cam_simplified} in a self referencing configuration where the carrier field is purely Gaussian. A similar approach could be used for external referencing configurations and/or other types of heterodyne imaging systems, although this presents other difficulties: the shape of the external reference must be known so that it's effects on the signal field measurement can be incorporated into the network's training data, and any image artifacts inherent to the imaging system also need to be accounted for in the training data.
\paragraph{}
The invention of phase cameras came from the need to measure the spatial overlap of the carrier and sideband fields inside a gravitational wave interferometer \cite{goda_frequency-resolving_2004}. Sustaining an operating state sufficient to detect gravitational waves requires the use of precise automated control systems that serve to maintain the length of the cavities and the alignment of the mirrors within the interferometer. These error signals are derived from the beats of the carrier and sideband fields which are measured using single \& quad element photo-diodes \cite{fritschel_alignment_1998}. The presence of high order modes in the sideband fields or a poor spatial overlap with the carrier field can introduce offsets into the error signals which results in a decrease in performance of the control systems and eventual decrease in detector sensitivity. Next generation gravitational wave interferometers will utilize higher circulating optical power and squeezed light to reduce quantum noise \cite{and_advanced_2010,acernese_advanced_2014, barsotti_squeezed_2018}. Both of these upgrades place a greater importance on the automated control systems. Misalignment or mismatch of the squeezed field into the output mode cleaner cavity results in lower levels of injected squeezing \cite{oelker_squeezed_2014}. Higher circulating power increases thermal deformation in the optics which creates higher order modes and poor spatial overlap of the sideband fields. Point absorbers \cite{brooks_point_2021} located on the surfaces of the optics also have a drastic impact on detector sensitivity at higher circulating power through the generation of higher order modes. Phase cameras are uniquely suited wavefront sensors to diagnose these issues as they individually measure the wavefront of each sideband field, in comparison to the Hartmann sensors \cite{brooks_ultra-sensitive_2007} which can only measure the combined wavefront of the carrier and sidebands.
\paragraph{}
The utilization of phase cameras inside gravitational wave interferometers requires the development of techniques for processing and analyzing the images they produce. Correcting the high spatial frequency defects caused by mode mismatch, thermal deformations \& point absorbers inside the interferometer requires a fast method for sensing these issues. The widespread success of machine learning algorithms for fast image analysis in other areas of research makes them a promising candidate for such applications.
\begin{figure*}[ht!]
    \centering
    \includegraphics[width=\textwidth]{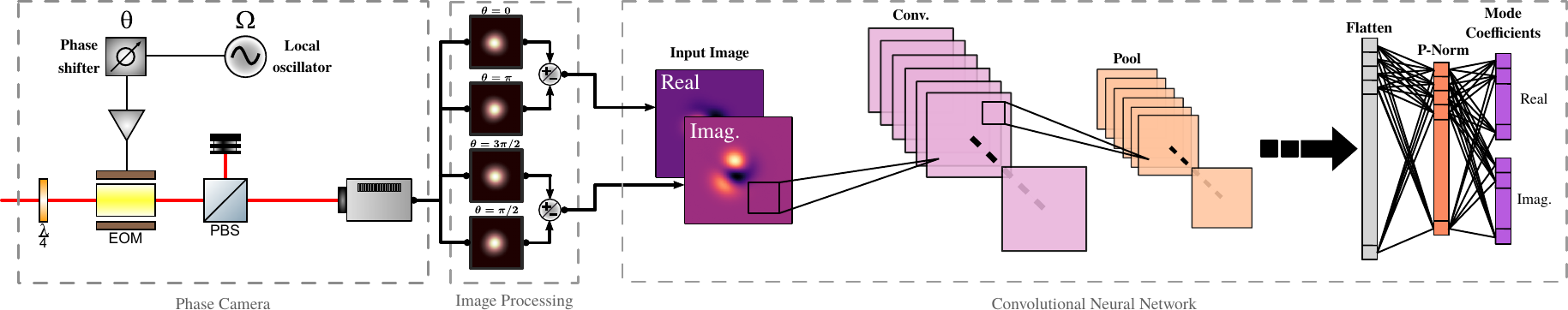}
    \caption{An overview of the convolutional neural network approach to perform mode decomposition of phase camera images.}
    \label{fig:network_decomp_pipeline}
\end{figure*}
\section{Overlap integral decomposition algorithm}
Mode coefficients were also calculated using a more traditional overlap integral method. This approach is not intended to represent the best possible (in accuracy or speed) integral based decomposition but rather for comparison, regardless it is described below:
\paragraph{}
The beam centre was calculated using a rudimentary centre-of-mass algorithm on the amplitude image map. The integral in Equation \ref{mode_coeff_integral_eqn} was evaluated using Simpson's method by generating HG modes with the desired $\Tilde{q}$ and centered around the calculated beam center. Since the phase camera images the beat of the signal and reference field, performing the overlap integral on the phase camera image yields a different set of coefficients to that of the signal field, $c^{'}_{nm}$:
\begin{equation}
    \label{primed_coeffs}
    \begin{split}
        c^{'}_{nm} = \int^{\infty}_{-\infty} \int^{\infty}_{-\infty} \Tilde{E}_{ref}^*(x,y,z_0)\Tilde{E}_{sig}(x,y,z_0)\\
        \times\text{HG}_{nm}^*(x,y,\Tilde{q}_{sig})\,dx\,dy
    \end{split}
\end{equation}
Where $\Tilde{E}_{ref}$ \& $\Tilde{E}_{sig}$ are the signal \& reference fields, evaluated at the imaging plane $z_0$. The desired coefficients of the signal field, $c_{nm}$, in the self referencing case with a purely Gaussian carrier field are related to these via the matrix equation below (see Appendix \hyperref[appendix_a]{A})
\begin{equation}\label{unprimed_coeffs}
\begin{gathered}
    c_{nm} = \int^{\infty}_{-\infty} \int^{\infty}_{-\infty} \Tilde{E}_{sig}(x,y,z_0)\text{HG}_{nm}^*(x,y,\Tilde{q}_{sig})\,dx\,dy \\
    c^{'}_{nm} = \gamma^{kl}_{nm} c_{kl} \\
    \begin{split}
        \gamma^{kl}_{nm} = \int^{\infty}_{-\infty} \int^{\infty}_{-\infty} \text{HG}_{00}^*(x,y,\Tilde{q}_{ref})\text{HG}_{kl}(x,y,\Tilde{q}_{sig})\\
        \times\text{HG}_{nm}^*(x,y,\Tilde{q}_{sig})\,dx\,dy
    \end{split}
\end{gathered}
\end{equation}
Each element of the matrix $\gamma$ is an integral of the product of three HG modes. This integral does not have a known analytical solution and must be evaluated numerically. Fortunately, $\gamma$ is symmetric under transpose, i.e $\gamma^{kl}_{nm}=\gamma^{nm}_{kl}$, which reduces the amount of computation needed. To calculate the mode coefficients $c_{nm}$, $c^{'}_{nm}$ are first calculated using an overlap integral and then the matrix $\gamma$ is created also using Simpson's method and then inverted to solve the matrix Equation \ref{unprimed_coeffs}.
\section{Convolutional neural network decomposition algorithm}

\subsection{Training data}
An overview of the network decomposition process is shown in Figure \ref{fig:network_decomp_pipeline}. Training a CNN requires a large amount of labelled training data. Here the training data is phase camera images labelled by corresponding mode coefficients. We elect to use artificially generated images due to the large amount needed and so that certain parameters of the training data can be tweaked easily.
\paragraph{}
The intensity images $\text{I}_\phi(x,y)$ taken by the camera at each demodulation phase $\phi$ can be emulated using Equation \ref{training_data}. The real and imaginary images of the beat field are then reconstructed by subtraction of the appropriate images.
\begin{equation}
    \label{training_data}
    \begin{aligned}
        & \text{I}_\phi(x,y) = \operatorname{Re} \Bigg[ \Tilde{E}_{ref}^*(x,y) \Tilde{E}_{sig}(x,y) e^{i\phi} \Bigg] \\
        & \text{Real} = \text{I}_{\phi=0}-\text{I}_{\phi=\pi}, \,\,\text{Imaginary} = \text{I}_{\phi=3\pi/2}-\text{I}_{\phi=\pi/2}
    \end{aligned}
\end{equation}
The desired output of the network for a given image are the mode coefficients $c_{nm}=r_{nm}e^{i\theta_{nm}}$ that describe the signal field $\Tilde{E}_{sig}(x,y)$. Calculating these requires the reference field $\Tilde{E}_{ref}(x,y)$ to be known. Additionally, the decomposition basis $\Tilde{q_{sig}}$ and maximum mode order must be specified. This means that these parameters must be passed to the network alongside the images or alternatively be fixed in the training data. We choose the latter as this greatly reduces the parameter space of the training data and required size of the network. A consequence of this is that the network must be retrained if either the reference field or decomposition basis changes. Fortunately retraining the network is significantly faster than training from scratch. Typically the reference field and basis would not change in a deployed phase camera application. Table \ref{table:training_data_parameters} shows the parameters used to in Equation \ref{training_data} to generate the training data.
\begin{table}[h]
\centering
\caption{Parameters used in Equation \ref{training_data} to define the network training dataset.}
\label{table:training_data_parameters}
\begin{tabular}{p{0.17\linewidth}  p{0.13\linewidth} p{0.5\linewidth}}
\hline
     Parameter & Fixed/ Variable & Value\\
     \hline
     $\Tilde{E}_{ref}(x,y)$ & fixed & \text{HG}$_{00}(x{-x_0}{-x_r},y{-y_0}{-y_r},\Tilde{q}_{ref})$\\
     $\Tilde{E}_{sig}(x,y)$ & variable & $\sum_{nm}c_{nm}\text{HG}_{nm}(x{-x_0},y{-y_0},\Tilde{q}_{sig})$\\
     $\Tilde{q}_{ref}$ & fixed & $0.1282+0.4832i$\\
     $\Tilde{q}_{sig}$ & fixed & $0.1282+0.4832i$\\
     $x_0,y_0$  & variable & $\in[-0.6\omega_0,+0.6\omega_0]$\\
     $x_r,y_r$ & fixed & 0.00\\
     Max$(n{+m})$ & fixed & 3\\
     Resolution & fixed & 128x128\\
     \hline
\end{tabular}
\end{table}

Images were generated using a pure Gaussian reference field with the same origin and basis as the signal field. The beam center was randomly varied by changing $x_0, y_0$, and the signal field was randomly varied by generating different mode coefficients. For demonstrative purposes in this work we restrict the maximum mode order to $(n+m)\le3$, however in the next section we show this technique works at higher mode orders.
\paragraph{}
It was found that sampling the mode coefficients from a purely random distribution was not ideal. Instead half in a batch were generated randomly and the other half generated with random amounts of power only in select modes. Doing so resulted in faster training convergence and better performance on other testing sets generated using the \textsc{Finesse} \cite{finesse} interferometer modelling software. The justification for this being that beams generated by optical cavities consist of power in only specific resonant modes and hence represent a smaller subset of the set of all the possible mode coefficients. It then makes sense to generate more of these kinds of beams in the training data.

\subsection{Mode decomposition network}
\begin{figure}[h]
    \centering
    \includegraphics[width=\linewidth]{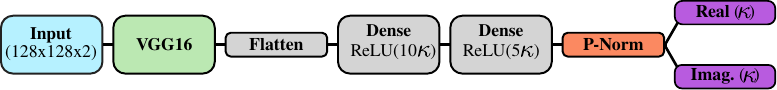}
    \caption{Overview of the decomposition network architecture, which contains a modified version of the VGG16 network \cite{simonyan_very_2015} with added dense and normalisation layers. The dense layers scale with the number of modes used in the decomposition \protect\scalebox{1.3}{$\kappa$}, and the custom P-Norm layer enforces the $\sum_{nm}|c_{nm}|^2=1$ normalisation condition.}
    \label{fig:network_architecture}
\end{figure}
We require a network architecture which can handle complex valued inputs and outputs. Complex valued CNNs are a potential solution but at the time of writing are less common and underdeveloped compared to their real valued counterparts. Using traditional CNNs and separating the complex information into separate image channels is a more convenient option. This allows the use of CNN architectures which have been developed for colour images such as the VGG16 network \cite{simonyan_very_2015} for transfer learning. Using amplitude \& phase in each channel to represent complex values is undesirable due to issues with phase wrapping causing discontinuities in the loss gradient. Instead we separate the real/imaginary components of the input phase camera images and the output mode coefficients into separate channels. The pre-processing stage involved normalising by the maximum absolute pixel value, and stacking the real \& imaginary images depth-wise. An overview of the network architecture is shown in Figure \ref{fig:network_architecture}.
\paragraph{}
We utilize the popular VGG16 network \cite{simonyan_very_2015} as the base of the decomposition network. The weights of the base network were initialized with those trained on the ImageNet \cite{deng2009imagenet} dataset but were not frozen during training. Various other base networks were explored including ones that consisted of only a few convolution and pooling layers. Through brief hyperparameter searches it was found that using VGG16 as the base network gave the best performance, although the reason why is not known. The task of mode decomposition should seemingly not require networks as deep as those designed for ImageNet type object classification problems. Yet we found that attempts using layers of a few consecutive convolutional and pooling layers ineffective. Using a shallower network may allow for both faster image evaluation and network training times, and as such should be the subject of future work.
\paragraph{}
The output of the VGG16 network was flattened to a single dimension using a global max pooling layer. This was followed by two dense layers of 10 and 5 times the number of modes \scalebox{1.3}{$\kappa$} being used in the decomposition. For reference the number of modes in a k$^{th}$ order decomposition is \scalebox{1.3}{$\kappa$ }$=(\text{k}^2+3\text{k}+2)/2$. The last two layers consisted of a custom 'P-Norm' layer which enforced the $\sum_{nm}|c_{nm}|^2=1$ power normalisation of the predicted coefficients and the output layer which separated the real/imaginary components of the coefficients for each mode.
\subsection{Training summary}
It was found that a straight forward training approach was ineffective for mode decompositions above 4$^{th}$ order. Despite extensive efforts exploring various training hyperparameters the network consistently converged to sub-optimal local minima corresponding to random guess-like performance. This behaviour led to the use of a curriculum learning \cite{bengio_curriculum_2009} approach, wherein the training data is provided in a specific order of complexity rather than randomly. This technique has been shown in some cases to improve performance and learning speed by modifying the optimization landscape to favour gradient descent algorithms \cite{hacohen_power_2019}.
\begin{figure}[h]
    \centering
    \includegraphics[width=0.75\linewidth]{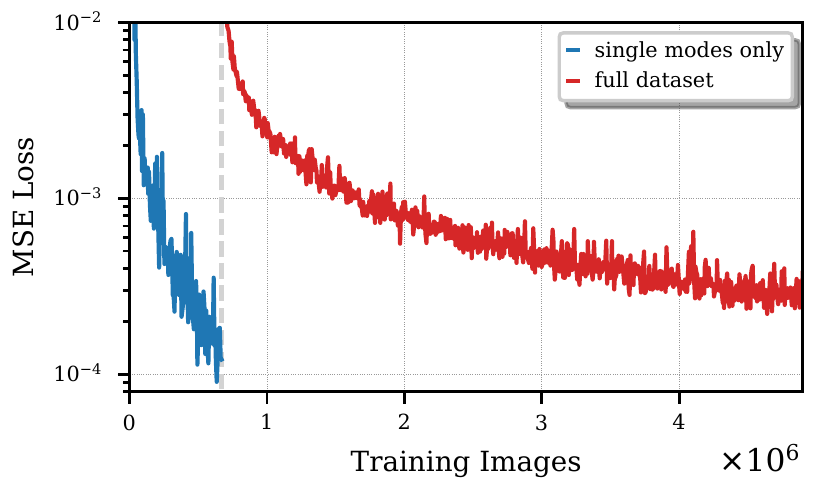}
    \caption{Summary of the network loss during the training process for the third order decomposition network. The first $\approx7\times10^5$ images consisted of only single modes.}
    \label{fig:training_summary}
\end{figure}
\begin{figure*}[ht!]
    \centering
    \includegraphics[width=\textwidth]{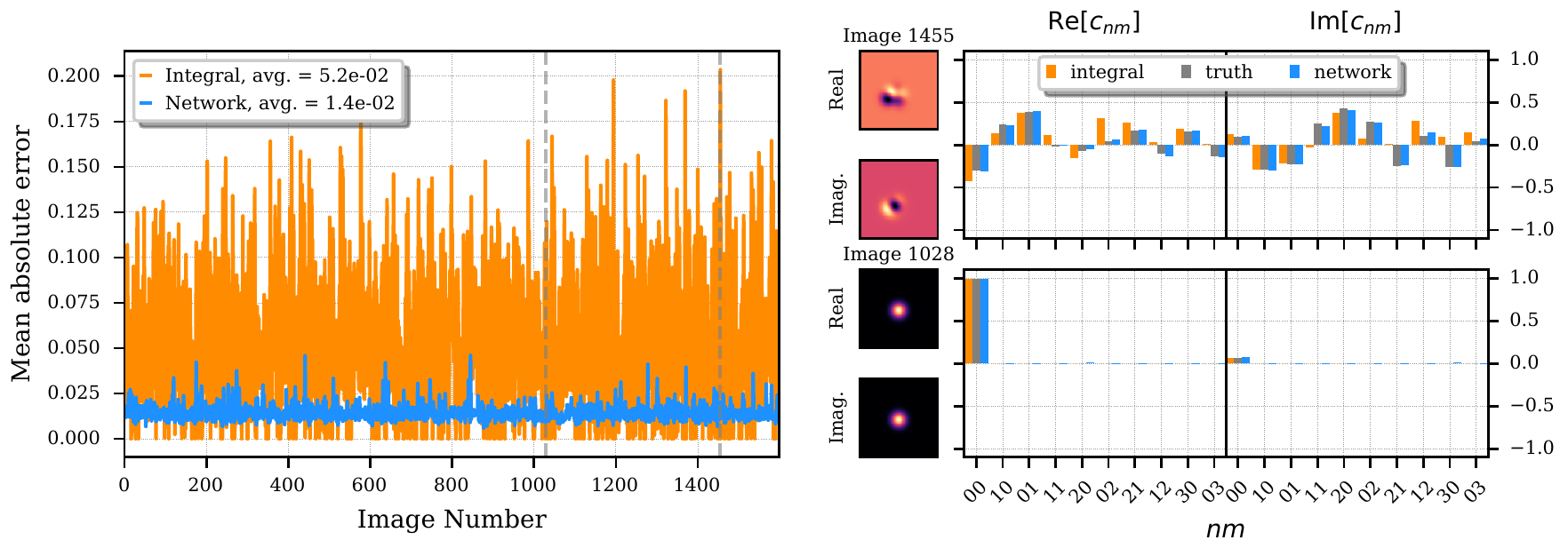}
    \caption{Results on the testing dataset for the network and integral decomposition. The mean absolute error (MAE) is defined in Equation \ref{eq:MAE}. Shown on the right are the best (bottom right) and worst (top right) test cases of the integral decomposition method. On average the neural network decomposition is $\approx 3.7$ times more accurate than the integral decomposition, however in cases where the beam has a spherically symmetric intensity distribution (bottom right) the integral method outperforms and has an error on the order of numerical precision. Conversely when the beam has an irregular intensity distribution the network greatly outperforms (top right).}
    \label{fig:testing_dataset_results}
\end{figure*}
\paragraph{}
The training consisted of two stages, the initial using training data that consisted of phase camera images containing only a single mode and the final using the complete training dataset. Both stages used the Adamax \cite{kingma_adam_2017} optimization algorithm, and learning rates of $10^{-4}$ \& $10^{-5}$ respectively. The initial stage lasted for $\approx7\times10^5$ images, and the final stage continued until the loss converged to an absolute delta of $< 10^{-5}$. The loss metric used was the squared error (MSE), shown in Equation \ref{eq:MSE}. We report in terms of numbers of images rather than epochs as the training dataset was not of a fixed size and instead generated on the fly in batches of 64 images. Figure \ref{fig:training_summary} shows the training summary for the 3$^{rd}$ order decomposition network, which converged after $\approx 5\times10^6$ total images. The same technique was used to train networks at higher orders, whose performance is reported later in Figure \ref{fig:higher_order_testing}.
\begin{equation}
    \label{eq:MSE}
    \text{MSE}=\frac{1}{\kappa}\sum_{nm}|c^{'}_{nm}-c_{nm}|^2
\end{equation}
\paragraph{}
The purpose of the initial stage was to better initialize the weights of the network for the final stage and avoid getting stuck in sub-optimal local minima. The intention was to force the network to learn to identify each of the individual modes first before seeing images with multiple modes. The filters learnt by the convolutional part of the network to decode the single mode images should be similar to those needed for the full dataset as each image is a simple linear combination of the individual modes. As such the weights of the network are initialized much closer to the optimal values, allowing the optimzation algorithm to avoid regions with sub-optimal local minima.
\paragraph{}
Training and testing was done using a moderately equipped workstation computer with a Ryzen 1700x 16 core CPU, 32 GB DDR4 RAM and a NVIDIA GTX-1070 GPU. The models were created and trained using the Keras \cite{chollet2015keras} machine learning API with a Tensorflow \cite{tensorflow2015-whitepaper} backend.
\section{Results}
\label{sec:testing_dataset_results}
\begin{figure}[h]
    \centering
    \includegraphics[width=0.75\linewidth]{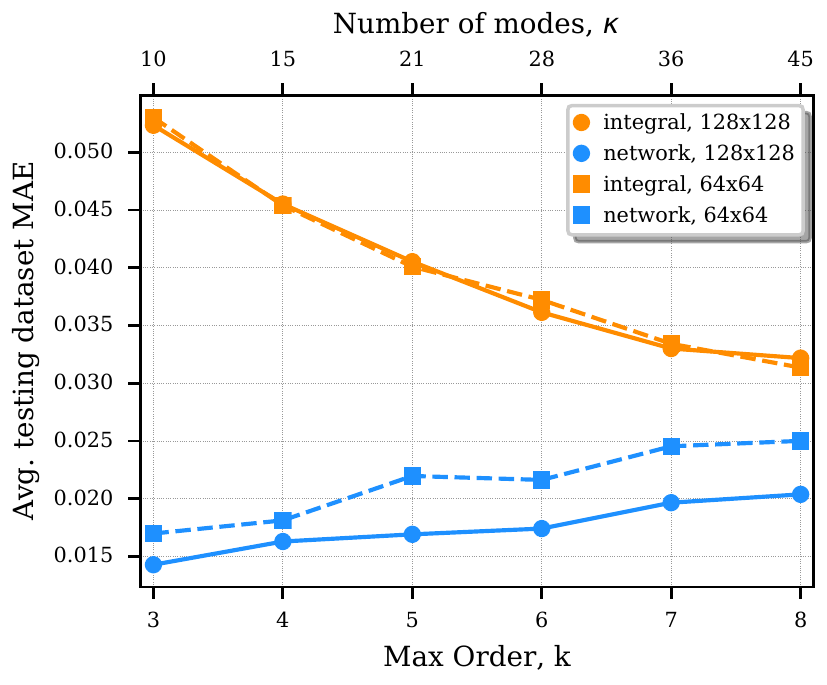}
    \caption{Average MAE over the testing dataset, using networks trained to perform the decomposition to higher orders and at image resolutions of 128x128 and 64x64 pixels.}
    \label{fig:higher_order_testing}
\end{figure}

A testing dataset of 1600 images was created in the same fashion as the training dataset, although importantly it consisted of images which the network had not explicitly seen during the training process. A comparison of the accuracy of the overlap integral algorithm and the network on this testing dataset is shown in Figure \ref{fig:testing_dataset_results}. Figure \ref{fig:higher_order_testing} shows the results on similar testing datasets created for networks trained to perform higher order decompositions and at a lower 64x64 image resolution. The mean absolute error (MAE) is defined in Equation \ref{eq:MAE}.
\begin{equation}
    \label{eq:MAE}
    \text{MAE}=\frac{1}{\kappa}\sum_{nm}|c^{'}_{nm}-c_{nm}|
\end{equation}
\paragraph{}
We see in Figure \ref{fig:testing_dataset_results} that the network based decomposition is on average $\approx 3.7$ times more accurate than the integral based decomposition over the testing dataset. The limiting factor for the integral method is the beam centering calculation, and the accuracy varies greatly across the testing dataset. Images where the integral method outperforms correspond to those with symmetric intensity distributions as these result in the center-of-mass being located at the beam center. The reverse is true for beams with an irregular intensity distribution. A similar testing dataset which had all beams centrally located and ignored the centering calculation was also created. The results of this showed that as expected the integral method had an accuracy limited only by the numerical precision related to the image resolution, but that the network decomposition had the same average error as on the non-centered beam testing dataset. This is evidence that the network has learnt a decomposition method which is not effected by the beam centering.
\paragraph{}
From Figure \ref{fig:higher_order_testing} there are two observed characteristics of the network decomposition - the average error for each mode increases with the number of modes present and there is a higher dependence on the image resolution than the integral based decomposition. The main error contribution in the integral method is from incorrect beam centering which mainly effects only a few modes. The MAE decreases at higher orders as this effect is averaged over a larger number of modes. The error from using a lower resolution has a comparatively much smaller effect which can not be seen. The network's higher dependence on image resolution is potentially due to the fact that it decomposes images through a series of learnt filters in the convolution layers. At lower image resolutions there may not be enough resolution inside the filter kernels to properly distinguish individual mode characteristics. The optical lock-in camera is resolution limited only by the camera used, but in this work hardware limitations prevented training neural networks to work at resolutions higher than 128x128.
\section{Conclusion}
We have trained a CNN to perform complete HG modal decomposition using a single image of a complex optical field taken by a phase camera. This machine learning decomposition scheme is the first to our knowledge to utilize images with complex phase information. Doing so allows the network to calculate the mode coefficients without a phase ambiguity, which is not possible with other schemes that use a single intensity image. We train our network using a curriculum learning technique that uses randomly generated phase camera images. This curriculum learning technique allowed the network to be trained to perform decomposition at higher orders than with a traditional learning approach. Our results showed that by training the network on non-centered beams it was able to learn a decomposition technique that was not effected by the beam centering. This allowed it to outperform a center-of-mass centering overlap integral decomposition algorithm on simulated datasets of images with non-centered beams.

\section*{Appendix A: Reference field unwrapping}
\label{appendix_a}
Heterodyne imaging techniques measure the beat of two frequency frequency separated fields, which we label the reference $\Tilde{E}_{ref}$ and signal $\Tilde{E}_{sig}$ fields. Demodulating at the frequency separation of these two fields at different phase quadratures allows the reconstruction of the beat field:
\begin{equation*}
    \Tilde{E}_{meas}(x,y,z_0) \propto \Tilde{E}^*_{ref}(x,y,z_0)\Tilde{E}_{sig}(x,y,z_0)
\end{equation*}
Calculating the mode coefficients of $\Tilde{E}_{sig}$ requires unwrapping the effects of the $\Tilde{E}_{ref}$ field which spatially envelopes the signal field. Calculating the mode coefficients of the measured field gives a set of mode coefficients, $c^{'}_{nm}$
\begin{equation*}
    \begin{split}
        c^{'}_{nm} = \int^{\infty}_{-\infty} \int^{\infty}_{-\infty} \Tilde{E}_{ref}^*(x,y,z_0)\Tilde{E}_{sig}(x,y,z_0)\\
        \times\text{HG}_{nm}^*(x,y,\Tilde{q}_{sig})\,dx\,dy
    \end{split}
\end{equation*}
Let the reference field be written as a sum of another set of mode coefficients $c^{''}_{ij}$, and similarly with the signal field.
\begin{equation*}
    \Tilde{E}_{ref} = \sum_{ij} c^{''}_{ij}\text{HG}_{ij}(x,y,\Tilde{q}_{ref})
\end{equation*}
\begin{equation*}
    \Tilde{E}_{sig} = \sum_{kl} c_{kl}\text{HG}_{kl}(x,y,\Tilde{q}_{ref})    
\end{equation*}
Adopting a notation where pairs of raised/lowered indices imply a summation we then have:
\begin{equation*}
\begin{split}
    c^{'}_{nm} = \int^{\infty}_{-\infty} \int^{\infty}_{-\infty} &
    c^{''}_{ij}{}^*\text{HG}^{*\, ij}(x,y,\Tilde{q}_{ref}) 
    \,c_{kl}\text{HG}^{kl}(x,y,\Tilde{q}_{sig}) \\
    & \times\text{HG}^{*}_{nm}(x,y,\Tilde{q}_{sig}) \,dx\,dy
\end{split}
\end{equation*}
placing the coefficients outside of the integral gives the matrix equation:
\begin{equation*}
    c^{'}_{nm} = c_{kl} \gamma^{kl}_{nm}
\end{equation*}
\begin{equation*}
\begin{split}    
        \gamma^{kl}_{nm} = c^{''}_{ij}{}^* \int^{\infty}_{-\infty} \int^{\infty}_{-\infty}& \text{HG}^{*\, ij}(x,y,\Tilde{q}_{ref}) 
        \text{HG}^{kl}(x,y,\Tilde{q}_{sig}) \\
        & \times\text{HG}^{*}_{nm}(x,y,\Tilde{q}_{sig}) \,dx\,dy
\end{split}
\end{equation*}
The signal field coefficients $c_{kl}$ can then be calculated by calculating the matrix $\gamma$ and inverting it to solve the above matrix equation.
\section*{Funding}
This project was funded by the Australian Research Council grant CE170100004.

\section*{Disclosures}
The authors declare no conflicts of interest.

\bibliography{sample}

\bibliographyfullrefs{sample}
\end{document}